\documentclass[prd,twocolumn,showpacs,amsmath,amssymb]{revtex4}

\usepackage{amssymb}
\usepackage{mathrsfs}
\usepackage{txfonts}

\usepackage{graphicx}
\usepackage{dcolumn}
\usepackage{bm}

\begin{document}

\title{Primordial non-Gaussianity in noncanonical warm inflation}

\author{Xiao-Min Zhang}
\email{zhangxm@mail.bnu.edu.cn}
\affiliation{Department of Physics, Beijing Normal University, Beijing 100875, China}
\author{Jian-Yang Zhu}
\thanks{Corresponding author}
\email{zhujy@bnu.edu.cn}
\affiliation{Department of Physics, Beijing Normal University, Beijing 100875, China}
\date{\today}
\begin{abstract}
We study the bispectrum of the primordial curvature perturbation on uniform-density hypersurfaces generated by a kind of the noncanonical warm inflation, wherein the inflation is provided by a noncanonical scalar inflaton field that is coupled to radiation through a thermal dissipation effect. We obtain an analytic form for the nonlinear parameter $f_{NL}$ that describes the non-Gaussianity in first-order cosmological perturbation theory and analyse the magnitude of this nonlinear parameter. We make a comparison between our result and those of the standard inflation and the canonical warm inflation. We also discuss when the contribution to the non-Gaussianity due to the second-order perturbation theory becomes more important and what effect can be observed. We take the Dirac-Born-Infeld (DBI) inflation as a concrete example to find how the sound speed and the thermal dissipation strength to decide the non-Gaussianity and to get a lower bound of the sound speed constrained by PLANCK.
\end{abstract}
\maketitle

\section{\label{sec1}Introduction}
Inflation, a quasi-exponential expansion of the very early Universe, is the most commonly adopted and successful model to explain the problems such as horizon, flatness and monopole \cite{Guth1981,Linde1982,Albrecht1982}. A most charming feature of the inflation is that it can predict the generation of a nearly flat spectrum of primordial perturbations, which fit the observations of cosmological microwave background (CMB) exactly. Such perturbations naturally arise from vacuum fluctuations of quantum fields in standard inflation \cite{LiddleLyth,Bassett2006} or thermal fluctuations in warm inflation \cite{BereraFang,Lisa2004,Berera2000}. There are two candidates of inflation so far - standard (also named cold inflation) and warm inflation as we just mentioned. They share the advantages such as successfully solving the horizon and flatness problem and naturally producing seed to give rise to the large scale structure of the Universe. In addition, warm inflation has some improved character compared to standard inflation. For example, the constant production of radiation through the interaction between inflaton and other sub-dominated fields can lead the Universe go into radiation dominated phase smoothly. And warm inflation can also cure the `eta problem' \cite{etaproblem} and the overlarge amplitude of the inflaton suffered in standard inflation \cite{Berera2005,BereraIanRamos}. The inflaton in warm inflation is not isolated as in cold inflation, so the evolution equation of inflaton is modified by an additional thermal damped term adding to it due to its interactions with other fields. So in warm inflation the damped effect is stronger and the slow-roll conditions are much more easily to be realized \cite{Ian2008,Campo2010,ZhangZhu}.

The inflaton usually used in standard inflation can be divided into two kinds: canonical and noncanonical scalar fields. Canonical field is the most commonly used one in inflationary scenario with the Lagrangian density $\mathcal{L}=X-V$, where $X=\frac12 g^{\mu\nu}\partial_{\mu}\phi\partial_{\nu}\phi$ is the kinetic term and $V$ is the potential of inflaton. However, noncanonical field with a general form of Lagrangian density $\mathcal{L}=\mathcal{L}(X,\phi)$ can generalize the scope of inflation, where $\mathcal{L}(X,\phi)$ is an arbitrary function of the inflaton $\phi$ and the kinetic term $X$ that satisfied the conditions proposed in \cite{Franche2010}. Much work have been done in the context of the noncanonical standard inflation \cite{refining2012,Mukhanov2006,Armendariz-Picon,Garriga1999,Gwyn2013,Tzirakis,Franche2010,Eassona2013,Bean2008} and a lot of novel phenomena have been revealed, such as the less restricted slow-roll conditions, the usual sub-light `sound speed' (denotes as $c_s$ ) for inflaton scalar perturbations, the modified expression of power spectrum and consistency equation, and in particular significant non-Gaussianity \cite{ChenHuang2007}. As far as we know, the warm inflation is always dealt with the canonical fields except in \cite{Cai2011} where a warm Dirac-Born-Infeld (DBI) inflationary model was proposed. Since the warm inflation and the noncanonical standard inflation both have some differentiable and good novel features, it's meaningful and interesting to combine them. We extend the warm inflation from canonical fields to a general noncanonical fields in \cite{Zhang2014} and build a new and broader picture of inflation. With stronger Hubble damping term and thermal damping term, stability analysis gave the even less restricted slow-roll conditions and a new but still scale-invariant power spectrum was obtained \cite{Zhang2014}. The energy scale of inflation when Hubble horizon crossing, the production of gravitational wave and consistency equation all acquire modification from both noncanonical effect and thermal effect. The non-Gaussianity generated in the new picture hasn't been fully researched and it's the main task of our paper.

Although the inflation is a successful and elegant picture to explain the problem of horizon, flatness and large scale structure of the Universe etc, there still exists a degeneracy problem \cite{Eassona2013}, i.e. a single set of observables maps to a range of different inflaton models (canonical and noncanonical fields with different potentials). Even a precise measurement of the spectral index, the running of spectral index, and the detection of gravitational wave will not allow us efficiently discriminate among them. There is another observable - non-Gaussianity can provide more information about the mechanism chosen by nature. Non-Gaussianity, as the name implies, is the deviation from a Gaussian statistics and the presence of higher-order correlation functions, such as the bispectrum \cite{Heavens1998,Ferreira1998} and the trispectrum \cite{Kunz2001}. The primordial curvature perturbation in slow-roll inflationary scenario is almost Gaussian with an almost scale-invariant spectrum. Research on the issue of non-Gaussianity in standard inflation has been done a lot in both canonical theory \cite{Gangui1994,Calzetta1995,David2005,Bartolo2004,Lyth2005,Zaballa2005} and noncanonical theory \cite{Creminelli2003,Tong2004}. The non-Gaussianities in multi-field inflation were calculated using $\delta N$-formalism in some papers \cite{Vernizzi2006,Battefeld2007,Tower2010}. The single-field slow-roll inflation model itself produces negligible non-Gaussianity, and in noncanonical or multi-field inflationary models, the non-Gaussianity can be significant. Non-Gaussianity was considered in different approach in \cite{MossXiong,Gupta2002,Gupta2006} in both strong and weak regime of canonical warm inflation. Our paper concentrates on the issue of non-Gaussianity in noncanonical warm inflation and make some comparisons between the new model and other inflationary models. The calculations about non-Gaussianity in kinds of models are almost all results of the first-order cosmological perturbation theory, and the analysis of the second-order cosmological perturbation theory is really complicated. Non-Gaussianity is often characterized by a nonlinear parameter $f_{NL}$. Some calculations in the context of the second-order perturbation theory were done \cite{Pyne1996,Komatsu2001,David2005,Lyth2005} and yield $f_{NL}\sim \mathcal{O} (1)$ regardless of the inflationary models. We'll also compare our result with the general second-order theory result. The new observations of PLANCK find no evidence for primordial non-Gausssianity for usual shapes (such as local, equilateral and orthogonal shapes) \cite{PLANCKNG}.

The paper is organized as follows: In Sec. \ref{sec2}, we introduce the new
noncanonical warm inflationary scenario and review the basic equations of the
new picture. Then in Sec. \ref{sec3}, we get the evolution equations for the first- and second-order scalar field perturbations. Then the nonlinear parameter $f_{NL}$ which can describe the non-Gaussian signatures in the new scenario is obtained in Sec. \ref{sec4} and the special DBI warm inflationary example is discussed. Finally, we draw the conclusions in Sec. \ref{sec5}.

\section{\label{sec2}noncanonical warm inflation dynamics}
Warm inflation occurs when there is a significant amount of radiation production during the inflationary epoch and the Universe is hot with a non-zero temperature $T$.
In noncanonical warm inflationary case, the total matter action of the multi-component Universe is
\begin{equation}\label{action}
  S=\int d^4x \sqrt{-g}  \left[ \mathcal{L}(X,\phi)+\mathcal{L}_R+\mathcal{L}_{int}\right],
\end{equation}
where the Lagrangian density of the noncanonical field is $\mathcal{L}_{non-can}= \mathcal{L}(X,\phi)$, $\mathcal{L}_R$ is the Lagrangian density of radiation fields and $\mathcal{L}_{int}$ denotes the interaction between the scalar fields. Usually the noncanonical Lagrangian should satisfy the conditions: $\mathcal{L}_{X}\geq0$ and $\mathcal{L}_{XX}\geq0$ \cite{Franche2010,Bean2008}, where a subscript $X$ denotes a derivative. The energy density and pressure of the noncanonical field are: $\rho(\phi,X)=2X\left(\partial\mathcal{L}/\partial X\right)-\mathcal{L}$, $p(\phi,X)=\mathcal{L}$. In the Friedmann-Robertson-Walker (FRW) Universe, the field is homogeneous, and we can get the equation of motion of the scalar field through varying the action. We consider the case that the Lagrangian density has a separable general form kinetic term and a potential term as in \cite{Zhang2014}, i.e. $ \mathcal{L}=K(X)-V(\phi)$, where $K$ is the noncanonical kinetic term which is weakly dependent or independent on $\phi$ \cite{Franche2010}.
Then the equation of motion of the noncanonical scalar field during warm inflation can be derived under some calculations \cite{Zhang2014}:
\begin{equation}\label{EOM1}
  \mathcal{L}_{X}c_{s}^{-2}\ddot{\phi}+(3H\mathcal{L}_{X}+\Gamma)\dot{\phi}+V_{eff,\phi}(\phi,T)=0,
\end{equation}
where $\Gamma$ is the dissipative coefficient due to thermal dissipation effect in warm inflation, $c_{s}^{2}=P_{X}/\rho_{X}=\left(1+2X\mathcal{L}_{XX}/\mathcal{L}_{X}\right)^{-1}$ is the `sound speed' which describes the traveling speed of scalar perturbations, $H$ is the Hubble parameter and $V_{eff}$ denotes the effective potential (the subscripts $\phi$ and $X$ both denote a derivative and for simplicity we'll write $V_{eff}$ as $V$ hereinafter). The interaction between the inflaton and other fields leads to thermal damping effect of inflaton and energy transfer to radiation \cite{BereraIanRamos}, which can be characterized by the entropy production equation:
\begin{equation}\label{entropy1}
    T\dot{s}+3HTs=\Gamma\dot{\phi}^{2}
\end{equation}
The number of e-folds is given by
\begin{equation}
N=\int H dt=\int\frac{H}{\dot{\phi}}d\phi\simeq-\frac{1}{M_p^2}\int_{\phi_{\ast}}
^{\phi_{end}}\frac{V(\mathcal{L}_X+r)}{V_{\phi}}d\phi,
\end{equation}
where $r=\Gamma/3H$ is the dissipation strength that parameterizes the efficiency of warm inflation ($r\gg1$ denotes strong warm inflation and $r\ll1$ denotes weak warm inflation).

Inflation is often associated with slow-roll approximation to drop the highest derivative terms in the equations of motion, the same in the new picture:
\begin{equation}\label{EOM2}
   (3H\mathcal{L}_{X}+\Gamma)\dot\phi+V_{\phi}(\phi,T)=0,
\end{equation}
\begin{equation}\label{entropy}
    3HTs-\Gamma \dot\phi^2=0.
\end{equation}
The validity of slow-roll approximation depends on slow-roll conditions characterized by slow-roll parameters defined as:
\begin{equation}
\epsilon =\frac{M_p^2}{2}\left(\frac{V_{\phi}}{V}\right) ^2, \eta =M_p^2\frac {V_{\phi \phi}}{V}, \beta
=M_p^2\frac{V_{\phi}\Gamma_{\phi}}{V\Gamma},
\end{equation}
and two parameters about the temperature dependence:
\begin{equation}
b=\frac {TV_{\phi T}}{V_{\phi}},c=\frac{T\Gamma_T}{\Gamma},
\end{equation}
where $M_p^2=1/8\pi G$ is the reduced squared Planck mass.
The slow-roll assumptions can be more easily to be guaranteed when
\begin{equation}
\epsilon\ll\frac{\mathcal{L}_{X}+r}{c^2_s},\beta\ll\frac{\mathcal{L}_{X}+r}{c^2_s},\eta\ll\frac{\mathcal{L}_{X}}{c^2_s}, b\ll\frac{min\{\mathcal{L}_{X},r\}}{(\mathcal{L}_{X}+r)c^2_s},|c|<4 \nonumber\\
\end{equation}
in noncanonical warm inflationary scenario \cite{Zhang2014}, which makes the analysis of the new model maneuverable.

\section{\label{sec3}evolution of perturbations during noncanonical warm inflation}
Thermal fluctuations dominate over quantum fluctuations in warm inflation and thermal noise is introduced in the evolution of the perturbations. Thermal fluctuations in radiation are coupled to the inflaton as a consequence
of the damping term in the inflaton evolution equation, and their amplitude is fixed
by a fluctuation-dissipation theorem. This means that both entropy and curvature
perturbations must be present. Fortunately on large scales
only the pure curvature perturbation survives \cite{Ian2008}. And primordial cosmological perturbations are usually expressed in terms of the curvature perturbation on uniform energy density hypersurfaces, denoted by $\zeta$. The quantity is widely used because it's conserved on large scales in simple models, even beyond linear order perturbation theory \cite{Bartolo2004,Lyth2005,Wands2003,Malik2004}. In order to treat the fluctuations of the primordial curvature perturbation, we utilize the similar stochastic approach which has been used in strong and weak warm inflation \cite{Gupta2002,Gupta2006}. In the linear order perturbation theory for the slow-roll single field inflation (noncanonical warm inflation is dominated by one single noncanonical inflaton field in overdamped slow-roll regime), the curvature perturbation on uniform density hypersurfaces is given by a gauge-invariant linear combination: $\zeta=\psi+\frac{H}{\dot\phi}\delta\phi$, where $\psi$ is spatial metric perturbation and $\delta\phi$ is perturbations about homogeneous inflaton fields. For convenience, we calculate in the commonly used spatially flat gauge (where $\psi=0$). Considering the perturbations, we expand the full inflaton field as
$\Phi(\mathbf{x},t)=\phi(t)+\delta\phi(\mathbf{x},t)$, with $\phi(t)$ being the homogeneous background field and $\delta\phi\ll \phi(t)$. The behaviour of a scalar field interacting with radiation can be analysed using the
Schwinger-Keldysh approach to non-equilibrium field theory \cite{Schwinger,Keldysh}. And the evolution equation of the full inflaton field coupled to radiation is described by a Langevin equation \cite{MossXiong,Chris2009}:
\begin{eqnarray}\label{pEOM1}
\mathcal{L}_{X}c_{s}^{-2}\ddot\Phi(\mathbf{x},t)+(3H\mathcal{L}_{X}+\Gamma)
\dot{\Phi}(\mathbf{x},t)+ V_{\phi}(\Phi(\mathbf{x},t))\nonumber\\-\mathcal{L}_{X}\frac{\nabla^2}{a^2}\Phi(\mathbf{x},t)=\xi(\mathbf{x},t),
\end{eqnarray}
where $\xi$ is the thermal stochastic noise in thermal system with zero mean $\langle\xi\rangle=0$. In
the high temperature limit $T\rightarrow\infty$, the noise source is Markovian: $\langle\xi(\mathbf{k},t)\xi(\mathbf{k'},t')\rangle=2\Gamma T(2\pi)^3\delta^3(\mathbf{k}-\mathbf{k'})\delta(t-t')$ \cite{Lisa2004,Gleiser1994}. Then we take the Fourier transform and obtain the evolution equation for the full inflaton in slow-roll regime:
\begin{equation}\label{EOMphik}
\frac{d\Phi(\mathbf{k},t)}{dt}=\frac{1}{3H\mathcal{L}_X+\Gamma}\left[-k^2\mathcal{L}_X\delta\phi(\mathbf{k},t)-V_{\phi}
(\Phi(\mathbf{k},t))+\xi(\mathbf{k},t)\right],
\end{equation}
where $\mathbf{k}$ is the physical momentum with the notation $\mathbf{k}\equiv\mathbf{k}_{phy}=\mathbf{k}_{com}/a$ ($\mathbf{k}_{com}$ is the comoving momentum) and the magnitude will be denoted as $k=|\mathbf{k}|$.

In order to calculate the predicted bispectrum from Eq. (\ref{pEOM1}), we should expand the inflaton fluctuations to second-order: $\delta\phi(\mathbf{x},t)=\delta\phi_1(\mathbf{x},t)+\delta\phi_2(\mathbf{x},t)$, where $\delta\phi_1=\mathcal{O}(\delta\phi)$ and $\delta\phi_2=\mathcal{O}(\delta\phi^2)$. Then the equations of motion for the first- and second-order fluctuations in Fourier space can be obtained from Eq. (\ref{EOMphik}):

\begin{eqnarray}\label{deltaphi1}
\frac{d}{dt}\delta\phi_1(\mathbf{k},t)&=&\frac{1}{3H\mathcal{L}_{X}+\Gamma}\left[-\mathcal{L}_Xk^2\delta\phi_1(\mathbf{k},t)\right.\nonumber\\
&-&\left.V_{\phi\phi}(\phi(t))\delta\phi_1(\mathbf{k},t)+\xi(\mathbf{k},t)\right],
\end{eqnarray}
\begin{eqnarray}\label{deltaphi2}
\frac{d}{dt}\delta\phi_2(\mathbf{k},t)&=&\frac{1}{3H\mathcal{L}_{X}+\Gamma}\left[-\mathcal{L}_Xk^2\delta\phi_2(\mathbf{k},t)-V_{\phi\phi}
(\phi(t))\delta\phi_2(\mathbf{k},t)\right.\nonumber\\ &-&\left.\frac12V_{\phi\phi\phi}(\phi(t))\int\frac{dp^3}{(2\pi)^3}\delta\phi_1(\mathbf{p},t)\delta\phi_1(\mathbf{k}
-\mathbf{p},t)\right. \nonumber\\ &-& \left. k^2\mathcal{L}_{XX}\int\frac{dp^3}{(2\pi)^3}\delta\phi_1(\mathbf{p},t)\delta X_1(\mathbf{k}
-\mathbf{p},t)\right],
\end{eqnarray}
where $\delta\phi_1$ is the linear response due to $\xi$ that acquires the whole thermal effect of the thermal noise and $\delta X_1(\mathbf{k})\simeq\dot\phi(t)\delta\dot\phi_1(\mathbf{k},t)$. From the stochastic statistics properties we can see that the first-order inflaton perturbations $\delta\phi_1$ are Gaussian fields and their bispectrum vanishes. The analytic form solutions to the evolution equations (\ref{deltaphi1}) and (\ref{deltaphi2}) are respectively:
\begin{eqnarray}\label{solution1}
\delta\phi_1(\mathbf{k},t)&=&A(k,t)\int_{t_0}^{t}dt'\frac{\xi(\mathbf{k},t')}{3H\mathcal{L}_X+\Gamma}A^{-1}(k,t')\nonumber \\&+&
A(k,t)\delta\phi_1(\mathbf{k}e^{-H(t-t_0)},t_0),
\end{eqnarray}
\begin{eqnarray}\label{solution2}
\delta\phi_2(\mathbf{k},t)&=&A(k,t)\int_{t_0}^{t}dt'\left[B(t')\int\frac{dp^3}{(2\pi)^3}\delta\phi_1(\mathbf{p},t')
\delta\phi_1(\mathbf{k}-\mathbf{p},t')\right.\nonumber\\ &-&\left. \frac{k^2\mathcal{L}_{XX}\sqrt{2X}}{(3H\mathcal{L}_X+\Gamma)^2}\int\frac{dp^3}{(2\pi)^3}\delta\phi_1(\mathbf{p},t')
\xi(\mathbf{k}-\mathbf{p},t')\right]\nonumber \\&\times &A^{-1}(k,t')+ A(k,t)\delta\phi_2(\mathbf{k}e^{-H(t-t_0)},t_0),
\end{eqnarray}
where the parameters $A$ and $B$ are given by
\begin{equation}\label{A}
A(k,t)=exp\left[-\int_{t_0}^{t}\left(\frac{\mathcal{L}_xk^2+V_{\phi\phi}}{3H\mathcal{L}_X+\Gamma}\right)dt'\right],
\end{equation}
\begin{equation}\label{B}
B(k,t)=-\frac{V_{\phi\phi\phi}}{6H(\mathcal{L}_X+r)}+\frac{k^2\mathcal{L}_{XX}\sqrt{2X}}{3H(\mathcal{L}_X+r)\tau}.
\end{equation}
In the solutions above, both the second terms on the right hand side are `memory' terms that reflect the state of the given mode at the beginning time $t_0$. And the parameter in the expression $A$ (denotes as $\tau(\phi)=\frac{3H\mathcal{L}_X+\Gamma}{\mathcal{L}_X^2k^2+m^2}$ where $m^2=V_{\phi\phi}$ is the effective squared inflaton mass) can describe the efficiency of the thermalizing process. We can see from Eq. (\ref{solution1}) that the larger squared magnitude of the physical momentum $k^2$ is, the faster the relaxation rate is. If $k^2$
 is sufficiently
large for the mode to relax within a Hubble time, then that mode thermalizes.
 While as soon as the physical momentum of a $\Phi(\mathbf{x},t)$ field mode becomes less than $k_F$ , it
essentially feels no effect of the thermal noise $\xi(\mathbf{k},t)$ during a Hubble time \cite{Berera2000,Zhang2014}. To quantify this criterion, the freeze-out momentum $k_F$ is defined by the condition
\begin{equation}\label{kf}
   \frac{\mathcal{L}_{X}\mathbf{k}_F^2+m^2}{(3H\mathcal{L}_{X}+\Gamma)H}=1
\end{equation}
That the mass term can be negligible is guaranteed by the slow-roll conditions and we can work out the freeze-out momentum $k_F=\sqrt{\frac{3H^2(\mathcal{L}_X+r)}{\mathcal{L}_X}}$. We can see that for strong regime of warm inflation, the fluctuations freeze in well before horizon exit, while in weak regime of warm inflation or the case that noncanonical effect dominate over thermal effect, fluctuations freeze in almost when horizon crossing.

\section{\label{sec4}non-Gaussianity for the noncanonical warm inflation}
For a perfect Gaussian distribution quantity, the two-point correlation function (or power spectrum as its Fourier transform) is all that is needed in order to completely characterize it from a statistical view and all the connected corrections higher than two-point vanish. If the distribution has some departure from perfect Gaussian one, then the higher-order correlation functions appear. The lowest-order statistics that is able to distinguish non-Gaussian from Gaussian perturbations is the three-point correlation function (or bispectrum). We know that the first-order inflaton perturbations $\delta\phi_1$ are Gaussian fields and their bispectrum must vanish. So the leading order of the non-Gaussianity of the fluctuations is the bispectrum generated from two first-order and one second-order fluctuations as
\begin{widetext}
\begin{eqnarray}\label{threepoint}
&&\langle\delta\phi(\mathbf{k}_1,t)\delta\phi(\mathbf{k}_2,t)\delta\phi(\mathbf{k}_3,t)\rangle \nonumber\\ &=&A(k_3,t)\int_{t_0}^t dt'A^{-1}(k_3,t')
\left[B(t')\int\frac{dp^3}{(2\pi)^3}\langle\delta\phi_1(\mathbf{k}_1,t)\delta\phi_1(\mathbf{k_2},t)\delta\phi_1(\mathbf{p},t')
\delta\phi_1(\mathbf{k}_3-\mathbf{p},t')\rangle\right.\nonumber\\ &-&\left. \frac{k^2\mathcal{L}_{XX}\sqrt{2X}}{(3H\mathcal{L}_X+\Gamma)^2}\int \frac{dp^3}{(2\pi)^3}\langle\delta\phi_1(\mathbf{k}_1,t)\delta\phi_1(\mathbf{k_2},t)\delta\phi_1(\mathbf{p},t')
\xi(\mathbf{k}_3-\mathbf{p},t')\rangle\right]+A(k_3,t) \langle\delta\phi_1(\mathbf{k}_1,t)\delta\phi_1(\mathbf{k_2},t)\delta\phi_2
(\mathbf{k}_3e^{-H(t-t_0)},t_0)\rangle\nonumber\\&+&(\mathbf{k}_1\leftrightarrow\mathbf{k}_3)+(\mathbf{k}_2\leftrightarrow\mathbf{k}_3).
\end{eqnarray}
\end{widetext}
The amplitude and slope of the bispectrum are determined at the time when cosmological scale exiting the horizon, nearly 60 e-folds before the end of inflation and for $\mathbf{k}_1$, $\mathbf{k}_2$ and $\mathbf{k}_3$ all within a few e-folds of crossing the horizon. The parameter $B(k,t)$ in the expression above is slowly varying so it can be approximated as a constant. When evaluating the three-point correlation function, the parameter $A(k,t)$ is weakly dependent on $t$ and can have the approximation $A(k,t-t_0)\approx1$ in the small time interval as stated in \cite{Gupta2002}. Since we have $k_F>H$ in noncanonical warm inflationary model, it implies
the $\delta\phi$ correlations that must be computed at time of Hubble horizon crossing $k=H$, are the thermalized correlations that were fixed at freeze-out time $k=k_F$ \cite{Berera2000}. So the time interval in the corrections is given by
\begin{equation}\label{time}
\Delta t_F=t_H-t_F\simeq\frac{1}{H}\ln\left(\frac{k_F}{H}\right).
\end{equation}
Thus Eq. (\ref{threepoint}) becomes
\begin{eqnarray}\label{threepoint1}
&&\langle \delta \phi ({\bf k}_1,t)\delta \phi ({\bf k}_2,t)\delta \phi (%
{\bf k}_3,t)\rangle \simeq 2B(t)\Delta t_F  \nonumber \\
&&\times \left[ \int \frac{dp^3}{(2\pi )^3}\langle \delta \phi _1({\bf k}%
_1,t)\delta \phi _1({\bf p},t)\rangle \langle \delta \phi _1({\bf k}%
_2,t)\delta \phi _1({\bf k}_3-{\bf p},t)\rangle \right.   \nonumber \\
&&+\left. ({\bf k}_1\leftrightarrow {\bf k}_3)+({\bf k}_2\leftrightarrow
{\bf k}_3)\right]
\end{eqnarray}
Now we'll dicuss the three-point statistics of the curvature perturbations. The bispectrum of curvature perturbation $\zeta$ is defined as
\begin{equation}\label{zetabispectrum}
\langle\zeta(\mathbf{k}_1)\zeta(\mathbf{k}_2)\zeta(\mathbf{k}_3)\rangle=(2\pi)^3B_{\zeta}(k_1,k_2,k_3)\delta^3(\mathbf{k}_1+\mathbf{k}_2
+\mathbf{k}_3).
\end{equation}
In spatially flat gauge we have the relation:
\begin{equation}\label{zeta}
\zeta=\frac{H}{\dot\phi}\delta\phi.
\end{equation}
 And since the inflaton fluctuation distributions are Gaussian dominating, the curvature perturbations are also Gaussian term dominating. For that case, the bispectrum can have the general form
\begin{equation}\label{bispectrum}
B_{\zeta}(k_1,k_2,k_3)=-\frac65f_{NL}\left[P_{\zeta}(k_1)P_{\zeta}(k_2)+cyclic\right],
\end{equation}
where $f_{NL}$ is the non-linear parameter that often used to describe the non-Gaussian signatures. Through the Eqs. (\ref{threepoint1}) (\ref{zeta}) and (\ref{bispectrum}), we can get the result:
\begin{eqnarray}\label{fnl}
f_{NL}&=&-\frac56\frac{\dot\phi}{H}2B(k_F,t_F)\Delta t_F\nonumber\\ &=&\frac56\frac{\dot\phi}{H}\frac{V_{\phi\phi\phi}(\phi(t_F))}{3H\mathcal{L}_X+\Gamma}\frac1H\ln\left(\frac{k_F}H\right)-\frac56
\ln\left(\frac{k_F}H\right)\left(\frac{1}{c_s^2}-1\right) \nonumber \\ &=&-\frac56
\ln\sqrt{\frac{3(\mathcal{L}_X+r)}{\mathcal{L}_X}}\left[\frac{\epsilon\varepsilon}{(\mathcal{L}_X+r)^2}+\left(\frac{1}{c_s^2}-1\right)\right],
\end{eqnarray}
where $\varepsilon=2M_p^2\frac{V_{\phi\phi\phi}}{V_{\phi}}$ can be seen as a slow-roll parameter which has the same magnitude as the slow-roll parameter $\eta$ in the monomial potential case. Thus we obtained the non-linear parameter $f_{NL}$ in the context of noncanonical warm inflationary theory which can be compared to the observations. The result we get in the new case can reduce to that in canonical warm inflationary case \cite{Gupta2002,Gupta2006} when $c_s=1$ and coincide with those in standard noncanonical case such as in DBI inflation and k-inflation qualitatively when noncanonical effect is dominated ($c_s\ll1$) \cite{ChenHuang2007,Tong2004,PLANCKNG}. The non-Gaussianity described by the non-linear parameter in strong regime of canonical warm inflation \cite{Gupta2002} has the same order as in standard inflation \cite{Gangui1994}. And the magnitude of non-linear parameter in weak regime of canonical warm inflation \cite{Gupta2006} is less than that of standard inflation for one order. The situation in our new case is more complicated for the dependence of more parameters. From the expression of Eq. (\ref{fnl}), and the analysis in \cite{Zhang2014,Gupta2002,Gupta2006} we can see that the non-Gaussianity in our new case is significant to some degree than in canonical warm inflation and the same order as in standard noncanonical inflation. In order to analysis the magnitude of $f_{NL}$, we utilize the slow-roll conditions we get in \cite{Zhang2014} to find that the first term in Eq. (\ref{fnl}) is really small when horizon crossing and thus we can have:
\begin{equation}\label{upperfnl}
|f_{NL}|\simeq\frac56\left(\frac{1}{c_s^2}-1\right)\ln\sqrt{\frac{3(\mathcal{L}_X+r)}{\mathcal{L}_X}}
\end{equation}
We find from the expression above that a small `sound speed' of the inflaton can much enhance the amount of the non-Gaussianity just as in noncanonical standard inflation, while the strong dissipative effect of warm inflation can also enhance the amount of the non-Gaussianity to some degree. The stronger the noncanonical effect is, the more the departure from Gaussianity is. The above analysis are results in the context of first-order cosmological perturbation theory, and there is a contribution that comes from the evolution of the ubiquitous second-order cosmological perturbation after inflation. The second-order perturbation evolution is model independent and introduce a contribution $f_{NL}\sim \mathcal{O} (1)$. So in the limit of low `sound speed', the first-order result is dominated while in the canonical standard inflationary case the first-order contribution is negligible and overwhelmed by the second-order one. However, the new observations of PLANCK satellite suggest no evidence for primordial non-Gaussianity \cite{PLANCKNG} and place a upper bound $|f_{NL}|<\mathcal{O}(10^2)$, so the `sound speed' should not be too small.

\section{\label{sec41} An example: DBI inflation}
Here we concentrate on one kind of important and representative noncanonical inflationary case: Dirac-Born-Infeld (DBI) inflation as a concrete example. The warm DBI inflationary case was specially considered in \cite{Cai2011}, but the specialized issue of non-Gaussianity was absent. DBI inflation \cite{ChenHuang2007,Tong2004,ChenX2005} is motivated by brane inflationary models \cite{Dvali1999,Burgess2001,Kachru2003} in warped compactifications \cite{Greene2000,Giddings2002,KachruS}. The slow-roll conditions in DBI inflation can be much more easily satisfied than in canonical standard slow-roll inflation \cite{Zhang2014}. There are two kinds of DBI inflationary models: named ultraviolet (UV) model (the inflaton moves from the UV side of the warped space
to the IR side under the potential) and infrared (IR) model (the inflaton moves from the IR side of the warped space to the UV side under the potential) \cite{ChenHuang2007,ChenX2005}. However, the UV DBI model is already
at odds with observations \cite{PLANCKNG}, so we consider only the IR DBI model here. The potential of IR DBI inflationary model is given by
\begin{equation}\label{IRV}
  V=V_0-\frac12\beta H^2\phi^2,
\end{equation}
where $\beta\leq0.7$ is constrained by PLANCK's observations \cite{PLANCKNG}.

\begin{figure}[tbp]
\begin{center}
\includegraphics[clip,width=0.48\textwidth]{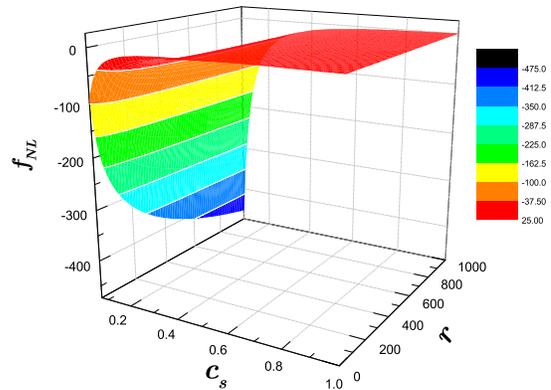}
\caption{(colour online).  The non-linear parameter $f_{NL}$ with different sound speed $c_s$ and dissipation strength $r$, $f_{NL}=f_{NL}(c_s,r)$ is shown.}
\label{Fig1}
\end{center}
\end{figure}

The DBI Lagrangian is of the form
\begin{equation}\label{DBIL}
\mathcal{L}_{DBI}=f^{-1}\left[1-\sqrt{1-2fX}\right]-V(\phi),
\end{equation}
where $\phi$ is the DBI inflaton field with potential $V$ and $f$ is the warp factor of an AdS-like throat. The kinetic term of DBI may depend on $\phi$ but the dependence is weakly \cite{Franche2010,Zhang2014} and so we consider the DBI Lagrangian with a constant warp factor $f=\Lambda^{-4}$ as in \cite{Franche2010} whose equation of motion can be given by Eq. (\ref{EOM1}). A general inflationary Lagrangian $\mathcal{L}(X,\phi)$ in a general gauge can reduce to canonical or DBI inflation in the specific gauge $\mathcal{L}_X=c_s^{-1}$. Thus the nonlinear parameter in our IR DBI warm inflationary example can reduce to
\begin{equation}\label{fnlDBI}
  f_{NL}=-\frac56\left(\frac1{c_s^2}-1\right)\ln\sqrt{3c_s\left(\frac1{c_s}+r\right)}.
\end{equation}
The picture of how the non-linear parameter $f_{NL}$ varies with sound speed $c_s$ and dissipative strength $r$ is shown in Fig. \ref{Fig1}.

Using the PLANCK observation constraint $f_{NL}^{DBI}=11\pm69$ \cite{PLANCKNG} we can obtain $c_s\geq0.09$ for our model, and in the standard DBI inflation the sound speed was constrained by $c_s\geq0.07$, so for completeness we plot the image in the interval $0.07\leq c_s\leq1$. And for completeness we plot the image in the interval $0\leq r\leq1000$ from weak regime ($r\ll1$) of warm inflation to strong regime ($r\gg1$) in Fig. \ref{Fig1}.

\begin{figure}[tbp]
\begin{center}
\includegraphics[clip,width=0.48\textwidth]{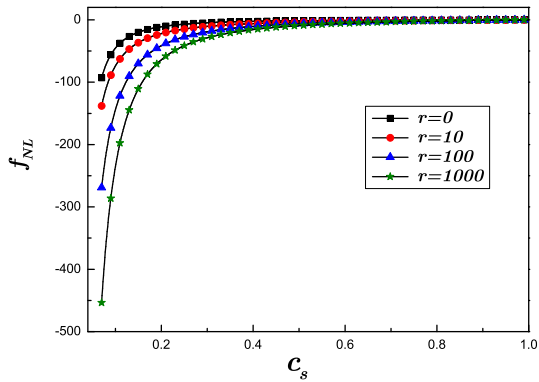}
\caption{(colour online).  The non-linear parameter $f_{NL}$ as a function of sound speed $c_s$ ($0.07\leq c_s\leq1$) in different dissipation strength of warm inflation.}
\label{Fig2}
\end{center}
\end{figure}

We can see from Fig. \ref{Fig1} that the non-Gaussian signature in noncanonical warm inflation is dominated by the noncanonical effect rather than the dissipation effect. From the Eq. (\ref{fnlDBI}) we can see that strong dissipation effect (i.e. large $r$) can also enhance the amount of non-Gaussianity, but it's not as obvious as small `sound speed'. To find intuitively how the non-linear parameter varies with different strength of warm inflation, we plot the picture as shown in Fig. \ref{Fig2}.
From Fig. \ref{Fig2} we can see that the larger $r$ is, the larger $|f_{NL}|$ is, especially in small `sound speed' regime, and when $c_s\rightarrow1$, the difference with different dissipation strength is not obvious. The larger the sound speed is, the weaker dependence of non-linear parameter on dissipation strength is.

\section{\label{sec5}conclusions}
We have introduced the noncanonical warm inflationary scenario proposed in \cite{Zhang2014} and reviewed the dynamics in the new scenario. The new model can be safely dealt with slow-roll approximation for the easily satisfied slow-roll conditions. In order to calculate the primordial non-Gaussianity of the curvature perturbation, we evolved the perturbations up to second-order in the spatially flat gauge. We obtained the equations of motion for the first- and second-order scalar field perturbations. Thermal stochastic noise was introduced in the evolution of fluctuations of scalar field as warm inflation indicated. We can get the analytic form solutions to the evolution equations. Based on these study, a non-zero bispectrum characterized by the non-linear parameter $f_{NL}$ of the curvature perturbation can eventually be obtained for the new model. The inflaton fluctuations is Gaussian term dominating and so the bispectrum can have the generic form of Eq. (\ref{bispectrum}). The non-Gaussian signature of the noncanonical warm inflation is more significant than that of the canonical warm inflation for the noncanonical effect especially in small sound speed regime. We find that both warm and standard inflation have almost the same magnitude for the non-Gaussianity ($f_{NL}\sim-(c_s^{-2}-1)$). The magnitude of non-Gaussianity in general noncanonical warm inflationary scenario was get by using the slow-roll conditions obtained in \cite{Zhang2014}. The magnitude was found to be much depended on the `sound speed' of the noncanonical inflaton scalar fluctuations. The `sound speed' can also describe the noncanonical effect of the inflaton, which is unity in canonical one and the smaller the `sound speed' is, the stronger the noncanonical effect is. The amount of non-Gaussianity can be enhanced by the larger departure from canonical case. The result we obtained is a first-order perturbation theory one and the primordial non-Gaussianity is dominated by first-order perturbations when the `sound speed' $c_s \ll1$ (the strong noncanonical effect in our case). In slow-roll canonical case, the general result $f_{NL}\sim \mathcal{O} (1)$ given by the model independent second-order perturbation theory can be important. Thus the second-order effect sets a lower bound of non-Gaussianity for inflation.

We give the DBI warm inflationary example to make the analysis of non-Gaussianity in our new scenario more visualized. That the PLANCK satellite find no evidence for primordial non-Gaussianity indicates that the `sound speed' should not be much less than unity. Using the PLANCK observations on primordial non-Gaussianity, we obtain the lower bound of `sound speed' in DBI warm inflationary model ($c_s\geq0.09$). We also give the dependence on thermal dissipative strength and reach the conclusion that it can enhance the non-Gaussianity but not very significant especially in weak noncanonical regime ($c_s\sim1$). The non-Gaussian signature in other noncanonical warm inflationary model such as tachyon warm inflation \cite{ZhangTachyon,Herrera2006} and power-law warm k-inflation also deserve much research and these can be our future work.

\acknowledgments This work was supported by the National Natural Science Foundation of China (Grants No. 11175019 and No. 11235003).

\end{document}